\input amstex
\documentstyle{amsppt}
{\catcode`\@=11\gdef\logo@{}}
\TagsOnRight
\loadbold
\pageheight{25 true cm}
\pagewidth{17 true cm}
\document
\centerline{\bf "Falling cat" connections and the momentum map}
\vskip 3cm
\centerline{Mari\'an Fecko ${}^{a)}$}
\centerline{Department of Theoretical Physics, Comenius University}
\centerline{Mlynsk\'a dolina F2, 842 15 Bratislava, Slovakia}
\vskip 3cm
{\bf Abstract}
\vskip 1cm
We consider a standard symplectic dynamics on $TM$ generated by a natural
Lagrangian $L$. The Lagrangian is assumed to be invariant with respect
to the action $TR_g$ of a Lie group $G$ lifted from the free and proper
action $R_g$ of $G$ on $M$. It is shown that under these conditions
a connection on principal bundle $\pi : M \rightarrow M/G$ can be
constructed based on the momentum map corresponding to the action $TR_g$.
A simple explicit formula for the connection form is given. For the
special case of the standard action of $G$ = SO(3) on
$M$ = ${\Bbb R}^3 \times \dots \times {\Bbb R}^3$ corresponding to a
rigid rotation of a N-particle system the formula obtained earlier
by Guichardet and Shapere and Wilczek is reproduced.
\vskip 5cm
PCAS : 02.40.Ma \ \   03.20.+i

\newpage

{\bf 1. Introduction}
\vskip 1cm
In their remarkable papers A.Guichardet [1] and A.Shapere and F.Wilczek
[2] pointed out that the phenomenon of reorientation of deformable bodies
(molecules represented by N point masses in [1] or cats, divers, astronauts etc.
in [2]) in space, for a long time to be known in the case of cats to
originate physically in the angular momentum conservation, lends itself
to a simple and powerful description within the framework of the theory
of connections (gauge structures). Namely they showed that in the
center-of-mass system (${\vec P} = {\vec 0}$) the condition of vanishing of
the total angular momentum (${\vec L} = {\vec 0}$) can be rephrased in terms
of the SO(3)-connection in the principle bundle $\pi : M \rightarrow M/SO(3)$,
where $M$ is the configuration space of the deformable body
(${\Bbb R}^{3N}$ minus some forbidden configurations in [1] or "the space
of located shapes" in [2]), where SO(3) acts by rigid rotations (without
deformation). In more detail the trajectories fulfilling ${\vec L} =
{\vec 0}$  represent the {\it horizontal} curves in the sense of the
connection ("vibrational curves" in [1] as opposed to purely rotational
ones given by (in general time dependent) rigid rotations).
\newline
\indent
In what follows we try to understand the origin of the connection
within the standard framework [3] of lagrangian mechanics on $TM$.
\newline
\indent
It is known that the central object providing the link between the
symmetries and conserved quantities in symplectic dynamics is the
{\it momentum map} [4,5]. Now both ${\vec P}$ and ${\vec L}$ result
(being linear in velocities) from the symmetries of rather special type,
namely those {\it lifted} to $TM$ from $M$. That is why the situation
under consideration is the following : we have a lagrangian system
($TM,L$) with appropriate action of a Lie group $G$ lifted from the
configuration space $M$. Then we show how one can construct (under
some restrictions on the Lagrangian $L$) a connection in the principal bundle
$\pi : M \rightarrow M/G$. This connection happens to coincide with
the one in [1,2] in the case treated there, i.e. for
$G$ = SO(3), $M$ being the configuration space of N-particle system.
\newline
\indent
The organization of the paper is the following. In Sec.2 (as well as in
Appendix A) the relevant facts concerning the momentum map within the
context mentioned above are collected. The construction of the connection
itself is described in Sec.3, the general properties of the latter are
discussed in
Sec.4. Several examples, including completely elementary ones as well as
the N-particle system are given in Sec.5. Some technicalities are treated in
appendices.
\vskip 1cm
{\bf 2. The momentum map for the lifted action $TR_g$}
\vskip 1cm
Let
$$R_g : M \rightarrow M \tag1$$
be a right action of a Lie group $G$ on a manifold $M$. Then the tangent
map
$$TR_g : TM \rightarrow TM $$
is a right action of $G$ on $TM$. Let $L : TM \rightarrow {\Bbb R}$ be
a $G$-invariant Lagrangian, i.e.
$$L \circ TR_g \equiv {(TR_g)}^* L = L \tag2$$
for all $g \in G$. The (exact) symplectic form on $TM$ is given by
([3]; see Appendix A)
$${\omega}_L = d {\theta}_L = d S(dL) $$
where (1,1)-type tensor field $S$ on $TM$ (almost tangent structure
$\equiv$ vertical endomorphism) is a lift of the identity tensor on
$M$ ($S = I^{\uparrow}$; in canonical local coordinates $x^i,v^i$ on
$TM$, $S = dx^i \otimes \frac{\partial}{\partial v^i}$ or
$S = dx^i \otimes \frac{\partial}{\partial {\dot x^i}}$ if the notation
$v^i \equiv {\dot x^i}$ is used).
Since ${\omega}_L$ is to be maximum rank 2-form, the condition
$$det ( \frac{{\partial}^2L}{\partial v^i \partial v^j}) \neq 0 $$
must be fulfilled (nondegenerate Lagrangian).
\newline
\indent
Let $a \in {\Cal G}$ (the Lie algebra of $G$), $X_a$ the corresponding
fundamental field of the action $R_g$ on $M$. Then the fundamental field
of the lifted action $TR_g$ is the {\it complete lift} ${\tilde X}_a$
(in coordinates  if $V = V^i {\partial}_i$ on $M$ then ${\tilde V} =
V^i {\partial}_i + {V^i,}_j v^j \frac{\partial}{\partial v^i}$ on $TM$).
Now
$${\Cal L}_{ {\tilde X}_a} {\theta}_L =({\Cal L}_{ {\tilde X}_a} S)
   (dL) + S(d {\tilde X}_a L) = {\theta}_{ {\tilde X}_a L} $$
(${\Cal L}_{\tilde V} S = 0 $ for any $V$). In the case of invariant
Lagrangian (3) gives
$${\tilde X}_a L = 0 \tag3$$
i.e.
$${\Cal L}_{ {\tilde X}_a} {\theta}_L =0 $$
Then
$$i_{ {\tilde X}_a} d{\theta}_L + d i_{ {\tilde X}_a} {\theta}_L
  = 0 $$
or
$$i_{\tilde X_a} {\omega}_L = - d P_a $$
($\Rightarrow {\tilde X}_a$ is hamiltonian field generated by $P_a$)
where $P_a : TM \rightarrow {\Bbb R}$ is defined by
$$P_a := <{\theta}_L, {\tilde X}_a> = S(dL,{\tilde X}_a) =
  S({\tilde X}_a) L = X^{\uparrow}_a L \tag4$$
($X^{\uparrow}_a$ is a {\it vertical lift} of $X_a$). Since $P_a$
depends linearly on $a \in {\Cal G}$, the {\it momentum map} associated
with the (exact symplectic) action $TR_g$ on $TM$
$$P : TM \rightarrow {\Cal G}^* $$
can be introduced by
$${<P(v),a>}_0 := P_a (v) \ \ v \in TM $$
where ${<.,.>}_0$ is the evaluation map (canonical pairing) for ${\Cal G}$
and its dual ${\Cal G}^*$. Fixing a basis $E_{\alpha}, {\alpha} = 1,\dots,
dim {\Cal G}$ in ${\Cal G}$ and the dual one $E^{\alpha}$ in ${\Cal G}^*$
one can write
$$P = P_{\alpha} E^{\alpha} $$
where
$$P_{\alpha} \equiv P_{E_{\alpha}} : TM \rightarrow {\Bbb R} $$
are the components of $P$ with respect to $E^{\alpha}$.
\newline
\indent
One verifies easily the important (equivariance) property of $P$
$${(TR_g)}^* P = {Ad}_g^* P $$
or in components
$${(TR_g)}^* P_{\alpha} = {({Ad}_g^*)}^{\beta}_{\alpha} P_{\beta} $$
where $Ad_g^* : {\Cal G}^* \rightarrow {\Cal G}^*$ is the coadjoint
action of $G$ on ${\Cal G}^*$. If ${\Omega}^k ({\Cal M},\rho)$ denotes
the space of k-forms of type $\rho$ on $G$-space ${\Cal M}$ (i.e.
V-valued k-forms on ${\Cal M}$ obeying $R^*_g {\sigma} = \rho
(g^{-1}) \sigma$, $\rho$ being a representation of $G$ in $V$), we see
that
$$P \in {\Omega}^0 (TM,Ad^*) \tag5$$
- it is ${\Cal G}^*$ - valued 0-form of type $Ad^*$ on $TM$. Thus a right
action (1) of $G$ on $M$ which is a symmetry of a non-degenerate Lagrangian
L (in the sense of (2) ) leads automatically to the existence of (5).
\vskip 1cm
{\bf 3. The construction of a connection form}
\vskip 1cm
Let $R_g$ be the action (1). In order to obtain a principal $G$-bundle
$$\pi : M \rightarrow M/G \tag6$$
the action is to be in addition free (all isotropy groups trivial) and
proper (the map $(g,x) \mapsto (x,R_gx)$ is proper, i.e. inverse images
of compact sets are compact). A connection form on (6) is
$A \in {\Omega}^1(M,Ad)$ such that
$$<A,X_a> = a \tag7$$
holds for all $a \in {\Cal G}$. Thus $P \in {\Omega}^0(TM,Ad^*)$ is
available whereas we need $A \in {\Omega}^1(M,Ad)$. These two objects are
different, but fortunately "not too much" and one can quite easily
obtain some $A$ from $P$.
\newline
\indent
First there is a bijection  between 1-forms on $M$ and functions
on $TM$ "linear in velocities" , viz.
$$\sigma (v) := {<{\tilde \sigma},v>}_{{\pi}_M (v)} $$
($\sigma \in {\Omega}^0(TM), {\tilde \sigma} \in {\Omega}^1(M)$), or in
coordinates
$${\sigma}_i v^i \leftrightarrow {\sigma}_i dx^i $$
Then if our $P$ were linear in velocities, one could associate with it
${\tilde P} \in {\Omega}^1(M,Ad^*)$ by
$${<{\tilde P},v>}_{\pi_M (v)} := P(v) $$
(the fact that ${\tilde P}$ really remains to be $Ad^*$-type is easily
verified). The demand of linearity in velocities of $P_{\alpha}$
restricts the form of Lagrangian : according to (4)
$$P_{\alpha}(v) = X^{\uparrow}_{\alpha} L = X^i_{\alpha}(x)
  \frac{\partial L(x,v)}{\partial v^i} $$
If this is to be of the form $P_{\alpha i}(x)v^i$, the Lagrangian has
to be {\it natural}, i.e.
$$L(x,v) = \frac{1}{2} g_{ij}(x)v^i v^j - U(x) \tag8$$
(a standard Lagrangian for potential system with time-independent holonomic
constraints). Then explicitly
$$P_{\alpha}(v) = (X_{\alpha}^{\uparrow} L)(v) = X^i_{\alpha}(x)
  g_{ij}(x)v^j = {({\flat}_g X_{\alpha})}_i(x)v^i $$
and
$${\tilde P}_{\alpha} = {({\flat}_g X_{\alpha})}_i(x)dx^i =
   {\flat}_g X_{\alpha} $$
where ${\flat}_g$ is the "lowering index" operator (by means of the
metric tensor $g$ on $M$ given by the kinetic energy
term in $L$) from vector to covector fields (the metric tensor $g$
is denoted by the same letter as the group element $g \in G$; the
proper meaning of $g$ is, however, always clear from the context).
One also verifies that (see (3) and (8) )
$${\Cal L}_{X_a} g = 0 $$
i.e. $G$ acts on ($M,g$) as a group of isometries ($X_a$ are the
Killing vectors).
\newline
The next step is a "correction" of $Ad^*$-type to $Ad$-type (needed for $A$).
This can be done by composition with a map ${\hat h} : {\Cal G} \rightarrow
{\Cal G}^*$ induced by some Ad-invariant non-degenerate bilinear form $h$ on
${\Cal G}$ (see Appendix B). Then
$${\hat A} := {\hat h}^{-1} \circ {\tilde P}  \in {\Omega}^1 (M,Ad) $$
i.e. ${\hat A}$ is already type $Ad$ ${\Cal G}$-valued 1-form on $M$.
\newline
Finally one has to check whether (7) is fulfilled . We have
$$<{\hat A},X_{\alpha}> = {\hat h}^{-1} <{\tilde P},X_{\alpha}> =
  <{\tilde P}_{\beta},X_{\alpha}> {\hat h}^{-1} (E^{\beta}) =
  <{\flat}_g X_{\beta},X_{\alpha}> h^{\beta \gamma} E_{\gamma} =$$
$$= g(X_{\alpha},X_{\beta}) h^{\beta \gamma} E_{\gamma} =
  C^{\gamma}_{\alpha} (x) E_{\gamma} $$
where
$$C^{\gamma}_{\alpha} := g_{\alpha \beta}(x) h^{\beta \gamma} $$
$$g_{\alpha \beta}(x) := g(X_{\alpha}, X_{\beta}) \tag9$$
Thus
$$<{\hat A},X_a> = C(x)(a) $$
where
$$C(x) : {\Cal G} \rightarrow {\Cal G}, \ E_{\alpha} \mapsto
  C^{\beta}_{\alpha} E_{\beta} $$
According to Appendix C the ${\Cal G}$-valued 1-form
$$A:=C^{-1} \circ {\hat A}=C^{-1} \circ {\hat h}^{-1} \circ {\tilde P}
  \tag10$$
has already all the necessary properties of a connection form, i.e.
$$R_g^*A = Ad_{g^{-1}}A $$
$$<A,X_a> = a $$
and defines thus a connection on $\pi:M \rightarrow M/G$. Explicitly
we have
$$A = C^{-1} \circ {\hat h}^{-1} ( {\tilde P}_{\alpha} E^{\alpha}) =
   {\tilde P}_{\alpha} h^{\alpha \beta} C^{-1} (E^{\beta}) =$$
$$= {\tilde P}_{\alpha} ( h^{\alpha \beta} h_{\beta \mu} g^{\mu \nu} )
  E_{\nu} = ( g^{\alpha \beta} {\tilde P}_{\beta}) E_{\alpha}$$
where $g^{\alpha \beta}(x)$ is the inverse to $g_{\alpha \beta}(x)$ defined in
(9). Thus it turns out to be given by a surprisingly simple expression,
viz.
$$A = A^{\alpha} E_{\alpha} = (g^{\alpha \beta} {\tilde P}_{\beta})
  E_{\alpha} = g^{\alpha \beta}({\flat}_g X_{\beta}) E_{\alpha} \tag11$$
\newline
\noindent
Note : notice that the bilinear form $h_{\alpha \beta}$ was present on the
scene only temporarily and it dropped out from the resulting formula
(and thus one does not need it in fact for the construction of $A$).
\vskip 1cm
{\bf 4. Some properties of the connection given by A}
\vskip 1cm
Let $\gamma : {\Bbb R} \rightarrow M$ be a curve on $M$ representing some
motion of the system under consideration. What does it mean in physical
terms if it is purely horizontal (i.e. represents a parallel translation
in the sense of $A$) ? According to (11) we have
$$<A, {\dot \gamma}> = 0 \ \Rightarrow \ <{\tilde P}_{\alpha},{\dot
   \gamma}> = 0$$
or
$$P_{\alpha} ( {\hat \gamma}(t)) = 0$$
where ${\hat \gamma}$ is the natural lift of $\gamma$ to $TM$
( $(x^i(t), {\dot x^i}(t))$ in coordinates). Thus a horizontal curve is
such motion of the system that all conserved quantities $P_{\alpha}$
have all the time zero value (remember ${\vec P} = {\vec 0}$, ${\vec L}
= {\vec 0}$ in the Sec.1.).
\newline
\indent
Now let $W \in Hor_xM$ be any horizontal vector. Then
$$0=<A,W> = g^{\alpha \beta}<{\tilde P}_{\beta}, W> E_{\alpha} =
  g^{\alpha \beta} g(X_{\beta},W) E_{\alpha}$$
or
$$g(X_{\alpha},W) = 0$$
for all $\alpha$. But $X_{\alpha}$ just span the vertical subspace
so that
$$Ver_xM \perp Hor_xM $$
Thus the horizontal subspace is simply the orthogonal complement of the
vertical one with respect to the scalar product in $T_xM$ given by the
kinetic energy metric tensor. Note that this serves as the {\it definition}
of the connection (it gives it uniquely) in [1] (in the special case
of $G = SO(3)$ etc. discussed in more detail further in Sec.5c.). In
the approach presented here it came as its {\it property}.
\vskip 1cm
{\bf 5. Examples}
\vskip 1cm
We illustrate the construction of the connection form $A$ on three examples,
the first two being completely elementary and the last one being that discussed
in [1] and [2].
\vskip 1cm
\noindent
{\it 5.1. A point mass on a board}
\vskip 1cm
\indent
Let us have a (one dimensional) board of mass $m_1$ laying on the surface
of the water  and denote $x$ the distance of its left end from some
reference point on the surface. Let $\xi$ denote the distance of a point
mass $m_2$ from the left end of the board. The Lagrangian of the system
reads
$$L(x,\xi,{\dot x},{\dot \xi}) = \frac{1}{2} m_1 {\dot x}^2 + \frac{1}{2}
  m_2 {({\dot x} + {\dot \xi})}^2 - U(\xi)$$
(interaction of the point mass $m_2$ with the board depends only on
their relative position). The translational invariance of the system means
that there is the action of $G \equiv {\Bbb R}$ on the configuration space
$M[x,\xi]$ given by
$$R_b : (x,\xi) \mapsto (x + b,\xi) \ \ b \in {\Bbb R} \equiv G$$
(the "unlocated shape" is given by the position of $m_2$ with respect
to the board, i.e. by $\xi$) such that $L$ is invariant with respect
to its lift
$$TR_b : (x,\xi,{\dot x},{\dot \xi}) \mapsto (x+b,\xi,{\dot x},
  {\dot \xi})$$
Now
$$X_1 = {\partial}_x \ \ \ g_{11} \equiv g(X_1,X_1) = m_1 + m_2$$
$$A = A^1 E_1 = g^{11} ({\flat}_g X_1) E_1 =
  \frac{1}{m_1 + m_2} ((m_1 + m_2)dx + m_2 d \xi ) =$$
$$=  dx + \frac{m_2}{m_1 + m_2} d\xi$$
(one can take $E_1 = 1$ since ${\Cal G} = {\Bbb R}$). The curve
$\gamma \leftrightarrow (x(t),\xi (t))$ is horizontal if
$<A, {\dot \gamma}> \equiv <A,{\dot x} {\partial}_x + {\dot \xi}
{\partial}_{\xi}> = 0$ , i.e. if
$${\dot x}(t) + \frac{m_2}{m_1 + m_2} {\dot \xi}(t) = 0$$
or
$$m_1 {\dot x}(t) + m_2 ({\dot x}(t) + {\dot \xi}(t)) = 0$$
which is just vanishing of the total (linear) momentum of the system.
\vskip 1cm
\noindent
{\it 5.2. A point mass on a gramophone disc}
\vskip 1cm
\indent
Let us have a gramophone disc (its moment of inertia with respect of
the axis being $I$) and a point mass $m$ on it. If the angle $\alpha$
measures the orientation of the disc with respect to the outer space
and $r,\varphi$ are the polar coordinates of the point mass $m$ with
respect to the disc, the Lagrangian of the system is
$$L(r,\varphi,\alpha,{\dot r},{\dot \varphi},{\dot \alpha}) =
  \frac{1}{2} I {\dot \alpha}^2 + \frac{1}{2} m ({\dot r}^2 + r^2
  {({\dot \alpha} + {\dot \varphi})}^2) - U(r,\varphi)$$
(interaction of the point mass $m$ with the disc depends only on
their relative position). The rotational invariance of the system means
that there is the action of $G \equiv SO(2)$ on the configuration space
$M[r,\varphi,\alpha]$ given by
$$R_{\beta} : (r,\varphi,\alpha) \mapsto (r,\varphi,\alpha + \beta)$$
(the "unlocated shape" is given by the position of $m$ with respect
to the disc, i.e. by $r,\varphi$) such that $L$ is invariant with respect
to its lift
$$TR_\beta : (r,\varphi,\alpha,{\dot r},{\dot \varphi},{\dot \alpha} ) \mapsto
 (r,\varphi,\alpha + \beta,{\dot r},{\dot \varphi},{\dot \alpha} )$$
Now
$$X_1 = {\partial}_{\alpha} \ \ \ g_{11} \equiv g(X_1,X_1) =
  I + m r^2$$
$$A = A^1 E_1 = g^{11} ({\flat}_g X_1) E_1 =
  \frac{1}{I + mr^2} ((I + mr^2)d\alpha + mr^2 d \varphi ) =
  d\alpha + \frac{mr^2}{I + mr^2} d\varphi$$
(one can take $E_1 = 1$ since ${\Cal G} = {\Bbb R}$ as in the previous
example). The curve
$\gamma \leftrightarrow (r(t),\varphi (t), \alpha (t))$ is horizontal if
$<A, {\dot \gamma}> \equiv <A,{\dot r} {\partial}_r + {\dot \varphi}
{\partial}_{\varphi} + {\dot \alpha} {\partial}_{\alpha} > = 0$ , i.e. if
$${\dot \alpha}(t) + \frac{mr^2}{I + mr^2} {\dot \varphi}(t) = 0$$
or
$$I {\dot \alpha}(t) + mr^2 ({\dot \alpha}(t) + {\dot \varphi}(t)) = 0$$
which is just vanishing of the total angular momentum of the system.
\newline
If $\sigma (t) \leftrightarrow (r(t),\varphi (t))$ is a curve in the space
of unlocated shapes $M/G$, the resulting curve in $M$ is
$\gamma (t) = {\sigma}^h (t) =$ the horizontal lift of $\sigma (t)$,
given by $(r(t),\varphi (t),\alpha (t))$, where
$$\alpha (t) = \alpha (0) + {\int}_0^t ( - {\dot \varphi}(s)
\frac{mr^2 (s)}{I + mr^2(s)} ) ds$$
In particular, the holonomy (the angle corresponding to the element
of SO(2) ) for the {\it closed} path (loop) $\sigma (0) = \sigma (1)$
is
$$\beta = \alpha (1) - \alpha (0) = - {\int}_0^1
\frac{mr^2 (s)}{I + mr^2(s)} ) {\dot \varphi}(s) ds$$
If for example the point goes round the disc once counterclockwise
at constant distance $r_0$ ($r(t) = r_0,\varphi (t) = 2\pi t$),
the net rotation of the disc is
$${\beta}_0 = -2\pi \frac{I_0}{I + I_0}  \hskip 1cm I_0 \equiv mr_0^2$$
(clockwise). Clearly $\alpha$ does not change
for radial motion (formally since $A^1_r = 0$).
\newline
There is nonzero curvature in this example being explicitly
$$F = DA = dA = d(\frac{mr^2}{I + mr^2}) \wedge d\varphi =
  (\frac{mr^2}{I + mr^2})'dr \wedge d\varphi \equiv
  \frac{1}{2} F^1_{r \varphi} dr \wedge d\varphi$$
\vskip 1cm
{\it 5.3. N-particle system}
\vskip 1cm
Let ${\vec r_a} \ , \ a = 1, \dots N$ denote the radius vector of a-th
particle , $x^i_a$ its i-th component (i = 1,2,3), $m_a$ its mass.
There is a natural action of the Euclidean group $G = E(3)$ on the
configuration space of the N-particle system, consisting in rigid
rotations and translations
$${\vec r_a} \mapsto {\vec r_a} B  + {\vec b} \ \ B \in SO(3)$$
We will treat the rotations and the translations separately. The standard
summation convention is adopted in what follows, i.e. the sum is
implicit for pairs of equal indices, otherwise the symbol of sum is written
explicitly.
\newline
The {\it translational} subgroup acts by
$$x^i_a \mapsto x^i_a + b^i$$
If ${\Cal E}_i$ is the standard basis of the Lie algebra ($\equiv
{\Bbb R}^3$), i.e. ${({\Cal E}_i)}^j = {\delta}^j_i$, then the
corresponding fundamental field is
$$X_i \equiv X_{{\Cal E}_i} = {\sum}_a {\partial}^a_i \equiv
  {(\vec {\nabla}_1} + \dots + {\vec {\nabla}_N )}_i$$
(${\partial}^a_i \equiv \frac{\partial}{\partial x^i_a}$).
The kinetic energy is
$$T = \frac{1}{2} {\sum}_a m_a {\dot x_a^k} {\dot x_a^k}$$
so that the metric tensor reads
$$g = {\sum}_a m_a dx_a^k \otimes dx_a^k$$
Then
$$g(X_i,X_j) = m {\delta}_{ij}$$
($m \equiv {\sum}_a m_a$ is the total mass). Since
$${\tilde P}_i = {\flat}_g X_i = m_a dx^i_a $$
we have the translational part of the connection
$$A_{tr} = A^i_{tr} {\Cal E}_i = \frac{1}{m} {\delta}_{ij} {\tilde P}_j
  {\Cal E}_i = \frac{m_a dx^i_a}{m} {\Cal E}_i$$
The {\it rotational} subgroup acts by
$$x^i_a \mapsto x^j_a B^i_j \hskip 5em  B \in SO(3)$$
If $E_i$ is the standard basis of the Lie algebra so(3),
i.e. ${(E_i)}_j^k =  - {\epsilon}_{ijk}$, the
corresponding fundamental field is
$$X_i \equiv X_{E_i} = - {\epsilon}_{ijk} x^j_a {\partial}^a_k \equiv
  - {({\vec r_a} \times {\vec {\nabla}_a})}_i$$
Then
$$g(X_i,X_j) = {\sum}_a ( {\delta}_{ij} {\vec r_a^2}-x^i_a x^j_a) = I_{ij}$$
where $I_{ij} ({\vec r_1}, \dots {\vec r_N})$ is the inertia tensor of the
configuration. Since
$${\tilde P}_i = {\flat}_g X_i = - {\epsilon}_{ijk} {\sum}_a m_a x^j_a dx^k_a
  \equiv - {({\sum} m_a {\vec r_a} \times d{\vec r_a})}_i$$
we have the rotational part (the one computed in [1,2]) of the connection
$$A_{rot} = A^i_{rot} E_i = I^{ij} {\tilde P}_j
   E_i = - I^{ij}({{\sum}_a m_a {\vec r_a} \times d{\vec r_a})}_j E_i$$
($I^{ij}$ being the inverse matrix to $I_{ij}$).
Putting both parts together the total (translational and rotational)
connection form reads
$$A = A_{tr} + A_{rot} =  \frac{m_a dx^i_a}{m} {\Cal E}_i +
   ( - I^{ij}({\vec r_1},\dots,{\vec r_N})
     {({\sum}_a m_a {\vec r_a} \times d{\vec r_a})}_j) E_i \equiv$$
$$\equiv \frac{{\tilde p}^i}{m} {\Cal E}_i - I^{ij} {\tilde L}_j E_i$$
(${\tilde p}^i, {\tilde L}_i$ being the total linear and angular momentum
1-forms respectively on $M$).
\newline
Let $\gamma (t) \leftrightarrow {\vec r_a}(t)$ be some motion of the
system, now. Then it is horizontal provided that $<A,{\dot \gamma}>
= 0$, i.e.
$$ \frac{m_a {\dot x^i_a}(t)}{m} {\Cal E}_i
    - {\sum}_a I^{ij}({\vec r_1}(t),\dots,{\vec r_N}(t)) m_a
     {({\vec r_a}(t) \times  {\dot {\vec r_a}}(t))}_j E_i = 0$$
or
$$m_a {\dot {\vec r_a}}(t) \equiv {\vec P}(t) = {\vec 0}$$
$${\sum}_a m_a {\vec r_a} \times {\dot {\vec r_a}} \equiv
  {\vec L} = {\vec 0}$$
Thus horizontal motion is such that the total (linear) momentum
${\vec P}$ as well as the total angular momentum ${\vec L}$ of the
 system vanish.
\vskip 1cm
{\bf 6. Conclusions and summary}
\vskip 1cm
In this paper we show that (under some restrictions mentioned in Sec.3.)
given a natural lagrangian system $(TM,L)$ with symmetry $G$ lifted from
the configuration space $M$ a connection in principle bundle
$\pi : M \rightarrow M/G$ can be constructed. The connection form $A$
is given by remarkably simple explicit formula $\thetag {11}$. It generalizes
"angular momentum equals zero" [6] connection from [1],[2], corresponding
to the group $G=SO(3)$. The construction of $A$ makes use of the momentum
map of the associated exact symplectic action of $G$ on $TM$, making the
link between the connection and conserved quantities explicit. A
calculation shows that the vertical and horizontal subspaces are mutually
orthogonal, which was used as the definition in [1].
\vskip 1cm
{\bf Appendix A : Some useful facts concerning the $TM$ geometry}
\vskip 1cm
Here we collect some more details on the constructions and objects on $TM$,
used in the main text (see [3]).
\newline
If $w \in T_xM$, its {\it vertical lift} to $v \in TM$ (${\pi}_M (v) = x$) is the
tangent vector in $t=0$ to the curve $t \mapsto v+tw$. The vector field (on
$TM$) obtained in such a way from the vector field $V$ on $M$ is denoted by
$V^{\uparrow}$. In canonical coordinates $(x^i,v^i)$ on $TM$
$$V \equiv V^i(x) {\partial}_i \mapsto V^{\uparrow} \equiv V^i(x)
  \frac{\partial}{\partial v^i}$$
Let $V$ be a vector field on $M$, and let us denote its local flow ${\Phi}_t$.
Then the generator
of the local flow $T{\Phi}_t$ on $TM$ is by definition the {\it complete lift}
${\tilde V}$ of $V$. In coordinates
$$V \equiv V^i(x) {\partial}_i \mapsto {\tilde V} \equiv V^i(x)
  \frac{\partial}{\partial x^i}+{V^i,}_j(x)v^j \frac{\partial}{\partial v^i}$$
If $w \in T_vTM$, then the map
$$S_v : T_vTM \rightarrow T_vTM \ w \mapsto {({\pi}_* w)}^{\uparrow}$$
(the lift being to $v$) is linear, giving rise to the (1,1)-tensor in $T_vTM$.
This pointwise construction defines a (1,1)-tensor field $S$ on $TM$
(almost tangent structure $\equiv$ vertical endomorphism), in coordinates
$S = dx^i \otimes \frac{\partial}{\partial v^i}$. Its properties used in
the main text are (easily verified in coordinates)
$${\Cal L}_{\tilde V} S = 0$$
$$S({\tilde V}) = V^{\uparrow}$$
If $B$ is (1,1)-tensor field on $M$, then its lift to $TM$ is defined
by $B^{\uparrow}(w) := {(B({\pi}_* w))}^{\uparrow}$. Then $S = I^{\uparrow}$
($I$ being the {\it unit} tensor field on $M$).
\vskip 1cm
{\bf Appendix B : The change of $Ad^*$ to $Ad$ via ${\hat h}^{-1}$}
\vskip 1cm
Let
$$h : {\Cal G} \times {\Cal G} \rightarrow {\Bbb R} $$
be non-degenerate bilinear form on ${\Cal G}$. It defines the map
$${\hat h} : {\Cal G} \rightarrow {\Cal G}^* $$
by ($a,b \in {\Cal G}$)
$$<{\hat h}(a),b>_0 := h(a,b) $$
($E_{\alpha} \mapsto h_{\alpha \beta} E^{\beta}$). If $h$ is Ad-invariant,
i.e.
$$h(Ad_g a, Ad_g b) = h(a,b), $$
then ${\hat h}$ satisfies
$$Ad^*_g \circ {\hat h} = {\hat h} \circ Ad_{g^{-1}} $$
Therefore
$$R_g^* ( {\hat h}^{-1} \circ {\tilde P}) = {\hat h}^{-1} \circ R_g^*
  {\tilde P} = {\hat h}^{-1} \circ Ad_g^* {\tilde P} =
  Ad_{g^{-1}} ({\hat h}^{-1} \circ {\tilde P}) $$
i.e. if ${\tilde P} \in {\Omega}^1 (M,Ad^*)$, then
${\hat A} \equiv {\hat h}^{-1} \circ {\tilde P} \in {\Omega}^1 (M,Ad)$.
\vskip 1cm
{\bf Appendix C : Transformation of the connection form into the
"canonical" form}
\vskip 1cm
Let $\pi : P \rightarrow M$ be a principal bundle and let ${\overline
A} \in {\Omega}^1 (P,Ad)$ define the connection by $Hor_p P := Ker
{\overline A}_p$. By definition $<{\overline A}_p,X_a> \in {\Cal G}$,
depending linearly on $a \in {\Cal G}$. Then
$$<{\overline A}_p,X_a> = C(p) (a) \tag C1$$
where
$$C(p) : {\Cal G} \rightarrow {\Cal G} $$
is invertible (lest some $X_a$ be horizontal). From
$$R_g^* {\overline A} = Ad_{g^{-1}} {\overline A} $$
and $\thetag {C1}$  one obtains
$$C(pg) = Ad_{g^{-1}} \circ C(p) \circ Ad_g $$
and therefore
$$A_p := C^{-1}(p) \circ {\overline A}_p \tag C2$$
has already the standard properties
$$R_g^* A = Ad_{g^{-1}} A$$
$$<A,X_a> = a \tag C3$$
This shows that although the standard requirement $\thetag {C3}$ on
connection form can be modified to a more general one $\thetag {C1}$,
it can be always simplified back to the "canonical" choice $\thetag {C3}$
via $\thetag {C2}$.
\vskip 1cm
{\bf 7. References}
\vskip 1cm
\noindent
$a)$ Present address : Department of Theoretical Physics, Co\-me\-nius
     University, Mlynsk\'a dolina F2, 842 15 Bratislava, Slovakia; e-mail:
     fecko\@fmph.uniba.sk  (published in J.Math.Phys. 36 (12) 6709-6719
     (1995))
\newline
\noindent
[1] A.Guichardet: "On rotation and vibration motions of molecules",
    Ann. Inst. Henri Poincar\'e, Vol.40, n.3, 1984, p.329-342
\newline
\noindent
[2] A.Shapere, F.Wilczek: "Gauge kinematics of deformable bodies",
    Am.J.Phys. 57 (6),514-518, June 1989
\newline
\noindent
[3] M.Crampin, F.A.E.Pirani : "Applicable Differential Geometry", Cambridge
    Univ. Press, Cambridge, 1987
\newline
\noindent
[4] J.M.Souriau : Structure des syst\`emes dynamiques, Dunod, Paris, 1970
\newline
\noindent
[5] V.I.Arnold : Mathematical Methods of Classical Mechanics,
    Benjamin/Cummings Reading MA, 1978, Appendix 5
\newline
\noindent
[6] R.Montgomery : "Isoholonomic Problems and Some Applications",
    Commun. Math. Phys. 128, 565-592 (1990)
\enddocument
\end